\documentclass[doublecol]{epl2}

\usepackage{graphicx}% Include figure files
\usepackage{dcolumn}% Align table columns on decimal point
\usepackage{amsmath}% Align table columns on decimal point
\usepackage{amssymb}% Align table columns on decimal point

\title{Stable crystalline lattices in two-dimensional binary mixtures of dipolar particles}
\author{Lahcen Assoud, Ren\'e Messina, Hartmut L\"owen}
\institute{
 Institut f\"ur Theoretische Physik II: Weiche Materie,
 Heinrich-Heine-Universit\"at D\"{u}sseldorf,
 Universit\"{a}tsstrasse 1,
 D-40225 D\"{u}sseldorf, Germany }

\pacs{82.70.Dd} {Colloids} %{First pacs description}
\pacs{61.50.Ah} {Theory of crystal structure, crystal symmetry; calculations and modeling} %{Second pacs description}
\pacs{61.66.Dk} {Alloys} %{Third pacs description}

\abstract{
The phase diagram of binary mixtures of particles
interacting via a pair potential of parallel dipoles is computed
at zero temperature as a function of composition and
the ratio of their magnetic susceptibilities. Using lattice sums,
a rich variety of different stable crystalline structures is identified including $A_mB_n$ 
structures. 
[$A$ $(B)$ particles correspond to large (small) dipolar moments.]
Their elementary cells consist of triangular, square, rectangular or rhombic
lattices of the $A$ particles with a basis comprising
various structures of $A$ and $B$
particles. For small (dipolar) asymmetry there are intermediate $AB_2$ 
and $A_2B$ crystals besides the pure $A$ and $B$ triangular crystals. 
These structures are detectable in experiments on granular
and colloidal matter.}

\begin{document}

\maketitle

While the freezing transition and the corresponding crystal lattice
in one-component systems is well-understood by now \cite{review,Dietrich_PRE_2001},
binary mixtures of two different particle species exhibits
a much richer possibility of different solid phases
For example, while a one-component hard sphere system
freezes into the close-packed face-centered-cubic
lattice \cite{Pronk}, binary hard sphere mixtures exhibit a huge variety
of close-packed structures depending on their diameter ratio.
These structures include $AB_{n}$ superlattices, where
$A$ are the large and $B$ the small spheres, with $n=1,2,5,6,13$. 
These structures were found in theoretical calculations \cite{Xu},
computer simulations \cite{Eldridge1,Eldridge2} and  in real-space experiments
on sterically stabilized colloidal suspensions \cite{Bartlett1,Bartlett2}. Much less
is known for soft repulsive interparticle interactions; most recent
studies on crystallization include attractions and consider
Lennard-Jones mixtures \cite{Ref27_SFB_Antrag_D1,Harrowell} or oppositely
charged colloidal particles \cite{Dijkstra1,Dijkstra2,Leunissen}.

In this letter we explore the phase diagram of a binary mixture interacting via 
a soft repulsive pair potential proportional to the inverse cube of the particle separation.
Using lattice sums, we obtain the zero-temperature phase diagram
as a function of composition and asymmetry, i.e. the ratio of the 
corresponding prefactors in the particle-particle interaction.
Our motivation to do so is threefold: 

i) First there is an urgent need to understand the effect of softness 
in general and in particular in two spatial dimensions. 
The case of hard interactions in two spatial dimensions, namely binary hard disks, 
has been obtained by Likos and Henley \cite{RefDA12} 
for a large range of diameter ratios.
A  complex phase behavior is encountered and it is unknown 
 how the phase behavior
is affected and controlled by soft interactions. 

ii) The model of dipolar particles considered in this letter is 
realized in quite different fields of physics.
Dipolar {\it colloidal particles} can be realized
by imposing a magnetic field \cite{Maret_1}.
In particular, our model is realized 
by  micron-sized superparamagnetic colloidal particles
which are confined to a planar water-air interface and
exposed to an external magnetic field parallel to the 
surface normal \cite{Maret_1,Maret_2,Maret_3,Koppl_PRL_2007,Magnot_PCCP_2004}. 
The magnetic field induces a magnetic dipole moment
on the particles whose magnitude is governed by the
magnetic susceptibility. Hence their interaction potential
scales like that between two parallel dipoles with the inverse
cube of the particle distance. Binary mixtures of colloidal particles
with different susceptibilities have been studied for
colloidal dynamics \cite{Naegele},
fluid clustering \cite{Hoffmann_PRL_2006,Hoffmann2},
and the glass transition \cite{Koenig}. 
A complementary way to obtain  dipolar colloidal particles
is a fast alternating electric field which generates effective dipole moments in the 
colloidal particles \cite{Blaaderen}. This set-up has been applied for two-dimensional binary
mixtures in Ref. \cite{Kusner}.
In {\it granular matter},
the model has been realized by mixing millimeter-sized steel and brass spheres \cite{Hay}
which are placed on a horizontal plate and exposed to a vertical magnetic field
such that the repulsive dipole-dipole interaction is the leading term.
Stable triangular $AB_2$ crystalline lattice were found \cite{Hay}. 
Layers  of {\it dusty plasmas}  involve particles whose interactions can be dominated
by that of dipoles \cite{Resendes,Shukla,Kalman}.
Other situations where two-dimensional mixtures of parallel dipoles are relevant 
concern {\it amphiphiles} (confined to a monomolecular film at an air-water interface
\cite{Seul}),  {\it binary monolayers} 
\cite{Israelachvili,Keller,Clarke}, 
{\it ferrofluid} monolayers \cite{Elias} exposed to a perpendicular magnetic field,
or thin films of {\it molecular} mixtures 
(e.g. of boron nitride and hydrocarbon molecules) with a large permanent dipole 
moment \cite{Wiechert}.
Hence, in principle, our results can be directly compared to various experiments
of (classical) dipolar particles with quite different size in quite different set-ups.

iii)  It is important to understand the different crystalline sub-structures in detail,
since a control of the  colloidal composite lattices
may lead to new optical band-gap materials (so-called photonic crystals) \cite{Pine}
to molecular-sieves \cite{Kecht} and to micro- and nano-filters 
with desired porosity \cite{Goedel}.
Nano-sieves and filters can be constructed on a colloidal 
monolayer confined at interfaces \cite{Goedel}. Their porosity is directly
coupled to their crystalline structure. For these applications, it is
mandatory to understand the different stable lattice types which
occur in binary mixtures. 

As a result, we find a variety of different stable composite lattices.
They include $A_mB_n$ structures with, for instance, $n=1,2,4,6$ for $m=1$. 
Their elementary cells consist of (equilateral) triangular, square, rectangular 
and  rhombic lattices of the $A$ particles. 
These are highly decorated by a basis involving either $B$
particles alone or both $B$ and $A$ particles. The topology of the 
resulting phase diagram differs qualitatively from that of hard disk mixtures \cite{RefDA12}.
For small (dipolar) asymmetries, for instance, we find intermediate $AB_2$ and $A_2B$
structures besides the pure triangular $A$ and $B$ lattices which are absent for hard disks.
Our calculations admit more candidate phases than considered
in earlier investigations \cite{RefDA20} where two-dimensional quasicrystals
were shown to be metastable. We further comment that we expect that colloidal glasses
in binary mixtures of magnetic colloids \cite{Koenig} are metastable as well but
need an enormous time to phase separate into their stable crystalline counterparts.

%%%%%%%%%%%%%%%%%%%%%%%%%%%%%%%%%%%%%%%%%%%%%%%%%%%%
%%%%%%%%%%%%%%%%     Our model    %%%%%%%%%%%%%%%%%%
%%%%%%%%%%%%%%%%%%%%%%%%%%%%%%%%%%%%%%%%%%%%%%%%%%%%
%
The model systems used in our study are binary mixtures of
dipolar  particles made up of 
two species denoted as $A$ and $B$. Each component $A$ and $B$  is characterized
by its dipole moment ${\mathbf m}_A$ and ${\mathbf m}_B$, respectively.
The particles are confined to a two-dimensional  plane and the dipole
moments are fixed in the direction
perpendicular to the plane. Thereby the
 dipole-dipole interaction is repulsive. Introducing the ratio $m=m_B/m_A$ of dipole
strengths $m_A$ and $m_B$, the pair interaction potentials between two $A$ dipoles, a $A$- and
$B$-dipole, and two $B$-dipoles at distance $r$ are
%
%%%%%%%%%%%%%%%%
\begin{eqnarray}
\label{eq_dipol}
V_{AA}(r)
\lefteqn{=V_0\varphi(r),
\quad V_{AB}(r)=V_0m\varphi(r),}\nonumber\\
& & V_{BB}(r)=V_0m^2\varphi(r),
\end{eqnarray} 
%%%%%%%%%%%%%%%%
%
respectively. 
The dimensionless function $\varphi(r)$ is equal $\ell^3/r^3$, where 
 $\ell$ stands for a unit length. The amplitude 
$V_0$
sets the energy scale. 

Our task is to find the stable crystalline structures adopted by the system at zero temperature.
We consider a parallelogram as a primitive cell which contains $n_A$ $A$-particles and $n_B$ $B$-particles.
This cell can be described geometrically by the two lattice vectors ${\mathbf a}=a(1,0)$
and ${\mathbf b}=a\gamma(\cos{\theta},\sin{\theta})$, where $\theta$ is the angle between 
${\mathbf a}$ and ${\mathbf b}$ and $\gamma$ is the aspect ratio ($\gamma=|{\mathbf b}|/|{\mathbf a}|$). 
The position of a particle $i$ (of species $A$)
and  that of a particle $j$ (of species $B$) in the parallelogram is specified by the vectors 
${\mathbf r}_{\rm i}^A=(x_i^{A},y_i^{A})$ and  
${\mathbf r}_{\rm j}^B=(x_j^{B},y_j^{B})$, respectively. 
The total internal energy (per primitive cell) $U$  has the form 
 
%%%%%%%%%%%%%%%%%%%%%%%%%%%%%%%%%%%
\begin{eqnarray}\label{eq_energy}
\lefteqn{U=\frac{1}{2}\sum_{J=A,B}
\sum_{i,j=1}^{n_J}\sideset{}{'}\sum_{\mathbf{R}}V_{JJ}
\left( \left| \mathbf{r}^J_i-\mathbf{r}^J_j+\mathbf{R} \right| \right)}
\nonumber\\
& & + \sum_{i=1}^{n_A}\sum_{j=1}^{n_B}\sum_{{\mathbf R}}
V_{AB}(\left| \mathbf{r}^A_i-\mathbf{r}^B_j+\mathbf{R} \right|),
\end{eqnarray} 
%%%%%%%%%%%%%%%%%%%%%%%%%%%%%%%%%%%
%
where ${\mathbf R}=k{\mathbf a}+l{\mathbf b}$ with $k$ and $l$ being integers.
The sums over  ${\mathbf R}$ in Eq. \ref{eq_energy}  run over all lattice cells where the prime indicates
that for  ${\mathbf R=0}$ the terms with $i=j$ are to be omitted. 
In order to handle efficiently the long-range nature of the dipole-dipole interaction,
we employed a Lekner-summation \cite{Lekner,Lekner_dip_2d}.

We choose to work at prescribed pressure $p$ and zero temperature ($T=0$). 
Hence, the corresponding thermodynamic potential is  the Gibbs free energy $G$. 
Additionally, we consider interacting dipoles at composition $X:=n_B/(n_A+n_B)$, 
so that the (intensive) Gibbs free energy $g$ per particle 
reads: $g=g(p,m,X)=G/(n_A+n_B)$.
At $T=0$,  $g$ is related to the internal 
energy per particle $u=U/(n_A+n_B)$ through $ g=u+p/\rho $, where the pressure $p$ is given by
$p=\rho^2(\partial u/\partial\rho)$, and $\rho=(n_A+n_B)/|{\mathbf a} \times {\mathbf b}|$ 
is the total particle density.
The Gibbs free energy per particle $g$ has been minimized with respect to 
$\gamma$, $\theta$  and the position of particles of species $A$ and $B$
within the primitive cell. 
To reduce the complexity of the energy landscape, we have  limited the number
of variables and considered the following candidates for our binary mixtures:
$A_4B$, $A_3B$, $A_2B$, $A_4B_2$, $A_3B_2$, $AB$,  $A_2B_2$, $A_2B_3$, $AB_2$, $A_2B_4$,
$AB_3$, $AB_4$ and $AB_6$.
For the $AB_6$ case we considered a triangular lattice formed  by the $A$ particles. 

%%%%%%%%%%%%%%%%%%%%%%%%%%%%%%%%
% Table 1
\begin{table}[ht]
%\begin{center}
\caption{
      The stable phases with their Bravais lattice and their basis.}
\begin{tabular}{l|l}
%\hline
\hline
Phase & Bravais lattice [basis]\\
\hline
&\\
{\bf T}($A$)          & Triangular  for $A$ [one $A$ particle]  \\

{\bf T}($B$)          & Triangular  for $B$ [one $B$ particle]  \\

{\bf S}($AB$)         & Square for $A$ and $B$ together \\ 
                      & [one $A$ and one $B$ particles] \\

{\bf S}$(A)B_n$       & Square for $A$ \\
                      & [one $A$ and $n$ $B$ particles]     \\

{\bf Re}$(A)A_mB_n$    & Rectangular for $A$ \\ 
                       &[$(m+1)$ $A$ and $n$ $B$ particles]         \\

{\bf Rh}$(A)A_mB_n$    & Rhombic for $A$ \\ & [$(m+1)$ $A$ and $n$ $B$ particles]         \\

{\bf P}$(A)AB_4$    & Parallelogram for $A$ \\ & [two $A$ and four $B$ particles]         \\

{\bf T}$(AB_2)$       & Triangular for $A$ and $B$ together \\
                & [one $A$ and two $B$ particles]      \\

{\bf T}$(A_2B)$       & Triangular for $A$ and $B$ together \\
                & [two $A$ and one $B$ particles]      \\

{\bf T}$(A)B_n$  & Triangular for $A$ \\ 
                & [one $A$ and $n$ $B$ particles]            \\ 
\hline
\label{tab.struct}
\end{tabular}
\end{table}
%%%%%%%%%%%%%%%%%%%%%%%%%%%%%%%%
%

\begin{figure*}[ht]
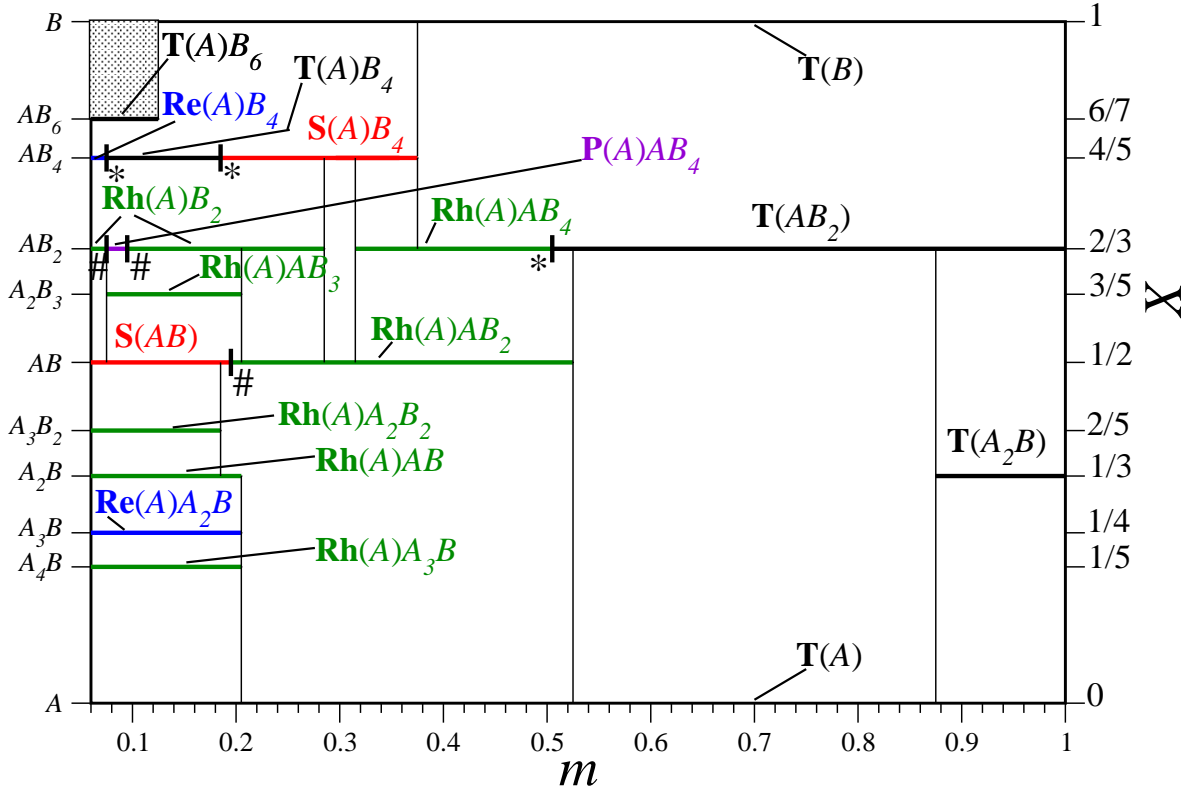

\onefigure[scale=0.6]{fig2.eps}
\caption{The phase diagram in the $(m,X)$ plane  of dipolar  asymmetry and composition 
         at $T=0$. The gray box denotes an unknown region.
         The symbol $\#$ ($\ast$) denote continuous (discontinuous) transitions.}
\label{fig.PD} 
\end{figure*}

The final phase diagram in the $(m,X)$-plane 
has been obtained by using the common tangent construction.
The dipole-strength ratio $m$ can vary between zero and unity.
A low value of $m$ (i.e., close to zero) corresponds to a large dipole-strength
asymmetry, whereas a high one (i.e., close to unity) indicates a weak
dipole-strength asymmetry.
%
% 
%
%%%%%%%%%%%%%%%%%%
% FIG 1
\begin{figure*}
\onefigure[width=16cm]{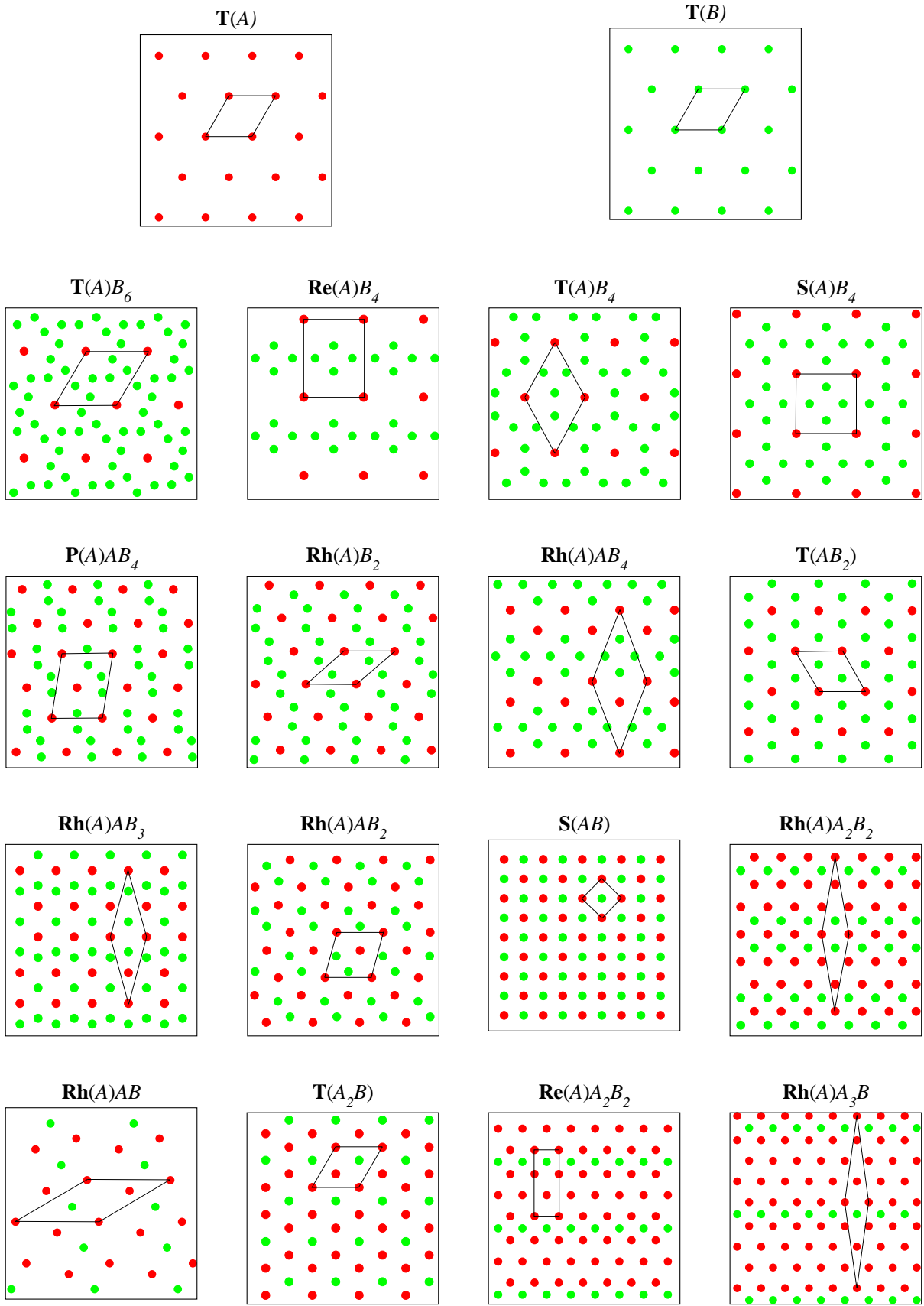}
\caption{The stable binary crystal structures and their primitive cells. The red dark (green light) 
         discs correspond to $A$ ($B$) particles.}
\label{fig.struct}
 \end{figure*}
%%%%%%%%%%%%%%%%%%
%
Our calculations show that all the  mixtures, except $AB_3$ and $A_4B_2$, are stable. 
Their corresponding crystalline lattices are depicted in figure \ref{fig.struct}
and the nomenclature is explained in Table \ref{tab.struct}. 
For the one component case [$X=0$ (pure $A$) and $X=1$ (pure $B$), see figure \ref{fig.PD}], 
we found an equilateral triangular lattice {\bf T}($A$) and {\bf T}($B$), respectively,  
as expected (see figure \ref{fig.struct}).

The most relevant and striking findings certainly concern the phase behavior
at weak dipole-strength asymmetry ($ 0.5 \lesssim m < 1$),  see figure \ref{fig.PD}.
Thereby, the only stable mixture $AB_2$ over such a large range of $m$
corresponds to the (``globally'' triangular) phase ${\bf T}(AB_2)$ 
(see figure \ref{fig.struct} and figure \ref{fig.PD}). 
This is in strong contrast to what occurs
with hard disk potentials \cite{RefDA12}, where $no$ mixture sets in at low size asymmetry.
At sufficiently low dipole-strength asymmetry ($m>0.88$), see figure \ref{fig.PD}, the mixture
$A_2B$, that also corresponds to a globally triangular crystalline structure
[namely ${\bf T}(A_2B)$, see figure \ref{fig.struct}],
is equally stable. 
The stability in the limit $m\to1$ of those globally triangular  
structures are fully consistent with the fact that one-component dipolar systems are triangular.

% 
%%%%%%%%%%%%%%%%%%
% FIG 2
%%%%%%%%%%%%%%%%%%

In the regime of strong dipole-strength asymmetry 
($ 0.06 < m  \lesssim 0.5$), see figure \ref{fig.PD},
the stability of the composition $X=1/2$, 
corresponding to the mixtures $AB$ and $A_2B_2$, 
is dominant and the phase diagram gets richer involving all the different structures
[except ${\bf T}(AB_2)$] shown in figure \ref{fig.struct}.
More specifically, for the composition  $X=1/2$, we have two phases
${\bf S} (AB)$ and ${\bf Rh}(A)AB_2$.   
The transition between these two phases is continuous as marked
by a symbol $\#$ in figure  \ref{fig.PD}.
For $X=2/3$, many stable phases emerge as depicted in figure \ref{fig.PD}.
%At even higher composition $X=4/5$,  three phases
%[T$(A)B_4$,  S$(A)B_4$ and R$(A)B_4$] are found. 
%The corresponding phase transitions are discontinuous
%(see the $\ast$ symbols in figure \ref{fig.PD}).  
In the $B$-rich region at large asymmetry, the stability
will involve many different structures which are probably not considered here.
Therefore we leave this region open, see the gray box in figure \ref{fig.PD}.
Below $X=2/3$, at large asymmetry ($m \lesssim 0.2$), the true phase diagram
will also involve a very dense spectrum of stable compositions,
as suggested by the already many stable compositions (see figure \ref{fig.PD}), 
which are not among the candidate structures considered here. 
This feature is very similar to the behavior reported 
in hard disk mixtures \cite{RefDA12}, where a {\it continuous} spectrum of 
stable mixtures is found for $X \leq 2/3$ at high size asymmetry.
In the limit $m \to 0$, a triangular lattice for the $A$ particles
will be stable with an increasingly complex substructure of $B$
particles.

In conclusion,  the ground-state phase diagram of a monolayer
of two-dimensional dipolar particles  shows a variety of different
stable solid lattices. The topology of the phase diagram is different from that of hard disks.
Whereas short-ranged interactions lead to a phase  separation into pure $A$ and $B$ crystals at low
asymmetries, there are two intermediate $A_2B$ and $AB_2$ mixtures for softer interactions.
This explains the experimental findings of 
Hay and coworkers \cite{Hay} who found an $AB_2$ crystal structure in
millimeter-sized steel and brass spheres \cite{Hay} which does not occur
to be stable for hard particles. A further more quantitative experimental 
confirmation of our theoretical predictions are conceivable either in 
suspension of magnetic colloids or for binary charged colloidal suspensions \cite{Palberg}
confined between two parallel glass plates \cite{Palberg2} or for any other situation
where two-dimensional dipolar particles are involved.

We finish with a couple of remarks: First, based on the present studies 
it would be interesting to study the behavior
of tilted dipoles where anisotropies and attraction play a significant role
\cite{Froltsov}. Our data may also serve as a benchmark  to perform further studies
on melting of the composite crystals and crystal nucleation out of the melt
in two spatial dimensions. 
The extension to one-component bilayers \cite{goldoni_prb,messina_prl} 
made up of dipolar particles would certainly be relevant. 
It would also be interesting to apply the method of
evolutionary algorithms \cite{Kahl} to the present problem in order to 
increase the basket of candidate phases.

We thank Christos Likos, Hans-Joachim Sch\"ope  and Thomas Palberg
for helpful discussions. 
This work
has been supported by the DFG within the SFB-TR6, Project Section D1.

\end{document}